\documentclass[conference]{IEEEtran}
\IEEEoverridecommandlockouts
\usepackage{cite}
\usepackage{amsmath,amssymb,amsfonts}
\usepackage{algorithmic}
\usepackage{graphicx}
\usepackage{textcomp}
\usepackage{xcolor}
\usepackage{braket}
\usepackage{caption}
\usepackage{subcaption}
\usepackage{tikz}
\usepackage{float}
\usepackage{booktabs}
\usepackage{comment}
\usetikzlibrary{calc} 
\usepackage{physics}
\usepackage{bm}
\def\BibTeX{{\rm B\kern-.05em{\sc i\kern-.025em b}\kern-.08em
    T\kern-.1667em\lower.7ex\hbox{E}\kern-.125emX}}
    \IEEEaftertitletext{\vspace{-1\baselineskip}}
\begin{document}

\title{Rydberg Atomic RF Sensor-based Quantum Radar%
\thanks{This paper has been accepted for presentation at the IEEE Applied Sensing Conference (APSCON 2026). This work was supported in part by the INSPIRE Faculty Fellowship (Reg. No.: IFA22-ENG 344), ANRF Prime Minister Early Career Research Grant (ANRF/ECRG/2024/000324/ENS), and the New Faculty Seed Grant from IIT Delhi.}

\thanks{Sourav Banerjee and Neel Kanth Kundu are with the Centre for Applied Research in Electronics, Indian Institute of Technology Delhi, India. Neel Kanth Kundu is also with the Bharti School of Telecom Technology and Management, Indian Institute of Technology Delhi, India (emails: souravbanerjer9@gmail.com, neelkanth@iitd.ac.in).}
}
\author{
\IEEEauthorblockN{Sourav Banerjee and Neel Kanth Kundu}
}


\maketitle
\begin{abstract}
Rydberg atom-based RF sensors offer distinct advantages over conventional dipole antennas for electric field detection. This paper presents a system model and performance analysis of a Rydberg atom-based quantum radar, which employs optical readout via lasers and photon detectors instead of circuit-based receivers. We derive the signal-to-noise ratio (SNR), compare it with classical radar, and estimate Doppler frequency using an invariant function-based method. Simulations show that the quantum radar achieves higher SNR and lower RMSE in velocity estimation than conventional radar.
\end{abstract}




\section{Introduction}
Conventional radar systems use a dipole antenna to detect an incoming electromagnetic (EM) echo signal that induces a current signal in the receiver circuit. Their receivers are based on the bulk properties of quantum mechanics, where the EM echo signal induces the flow of a large number of electrons in the receiver antenna circuit. Recently, Rydberg atom-based RF sensors have emerged as a promising technology that is fundamentally different from the conventional dipole antenna-based RF detectors \cite{anderson2020rydberg,gordon2019weak}. 

Most of the current works on Rydberg atom-based RF sensing have demonstrated the feasibility of measuring the amplitude, phase, polarization, and power of the electric field in a laboratory setup. Rydberg atom-based communications systems capable of detecting QPSK, AM, and FM modulation have also been experimentally demonstrated \cite{meyer2018digital,borowka2022sensitivity}. However, the system design and working principle of a Rydberg atom-based radar system have not yet been investigated in the literature. Prior works on quantum radars require microwave-entangled quantum states, which require dilution refrigerators \cite{sorelli2021detecting}. Furthermore, the joint detection at the receiver using the stored idler and reflected signal is practically challenging to implement \cite{sorelli2021detecting}. In order to overcome these challenges, we propose a quantum radar system that uses a conventional microwave source at the transmitter. However, the detector at the receiver is replaced with a Rydberg-atom-based RF sensor.

Since the detection principle of Rydberg atom-based receivers and conventional dipole antennas is fundamentally different, the subsequent radar signal processing also requires separate attention for Rydberg atom-based radar systems. In this paper, we address these open research problems. The main contributions of this paper can be summarized as follows:
\begin{enumerate}
    \item We present a system model for a modern quantum radar system using a Rydberg atom as the RF receiver.
   \item We analyze the SNR characteristics of the Rydberg atom-based radar system, and evaluate the target velocity estimation performance by employing invariant function based frequency estimation methods \cite{candan2021frequency}.  
   \item We compare the performance of the Rydberg atom quantum radar with that of the classical radar.
\end{enumerate}
\vspace{-3pt}

\section{Measurement of an Electric Field by Rydberg Atom-Based RF Sensor}
Rydberg atom-based electric field sensors are composed of alkali atoms (e.g., Cesium and Rubidium) in a glass cell of dimension a few centimeters in size \cite{anderson2020rydberg}. We consider a ladder-type four-level atom with states \(|1\rangle\) (ground state), \(|2\rangle\) (excited state), \(|3\rangle\)  and \(|4\rangle\) (Rydberg states). A weak probe laser (frequency \(\omega_p\)) drives the \(|1\rangle\leftrightarrow|2\rangle\) transition with Rabi frequency \(\Omega_p\), while a strong coupling laser (frequency \(\omega_c\)) drives \(|2\rangle\leftrightarrow|3\rangle\) with Rabi frequency \(\Omega_c\). An RF field of frequency \(f_{\rm RF}\) couples the Rydberg pair \(|3\rangle\leftrightarrow|4\rangle\) with RF Rabi frequency \(\Omega_{\rm RF}\).
The Rabi frequency for a field \(E\) coupling levels \(i\) and \(j\) is
\(\Omega_{ij}=\wp_{ij}E/\hbar\),
where \(\wp_{ij}\) is the transition dipole matrix element.
The transmitted probe power through a vapor cell (adiabatic approximation) is
\begin{equation}
P(t)=P_{\rm in}\exp\!\big[-k_p L\,\Im\{\chi(t)\}\big],
\label{eq:transmission}
\end{equation}
with \(P_{in}\) the input probe power intensity, \(k_p=2\pi/\lambda_p\) the probe wavenumber, \(L\) the cell length, and \(\chi(t)\) the complex susceptibility seen by the probe. The susceptibility is proportional to the density matrix element \(\rho_{21}\):
\begin{equation}
\chi(t)=C_0\,\rho_{21}(t),
\qquad \textit{with} \qquad
C_0 \equiv -\frac{2N_0\wp_{12}}{\epsilon_0\hbar\Omega_p},
\end{equation}
where \(N_0\) is the atomic density
\cite{FleischhauerEIT}. From \cite{chen2025harnessingrydbergatomicreceivers} we get, under resonant conditions ($\Delta_p = \Delta_{RF} = 0$, $\gamma_{41} = \gamma_{31}=0$), the imaginary part of $\rho_{21}$ is:
\begin{equation}
\Im(\rho_{21}) = -\Omega_p \gamma_{21}(\frac{\Omega_c^4}{8(\frac{\Omega_{RF}^2}{4\Delta_c}+\Delta_c)^2}+2\gamma_{21}^2)^{-1}
\end{equation}

For the sake of completeness, we need to provide a detailed discussion of the measurement of an incident RF signal using a local oscillator (LO). But for brevity, we refer to section III B of \cite{chen2025harnessingrydbergatomicreceivers}, a tabulated summary is provided in Table \ref{tab:rydberg_summary_compact}.
\begin{table*}[t]
\centering
\caption{\small Summary of LO-dressed Rydberg atomic receiver model (Section III.B of \cite{chen2025harnessingrydbergatomicreceivers}).}
\label{tab:rydberg_summary_compact}
\begin{tabular}{@{}ll|ll@{}}
\toprule
\textbf{Expression} & \textbf{Description} & \textbf{Expression} & \textbf{Description} \\ \midrule

$E_{\text{1}} = A_{\text{1}} \cos(2\pi f_{\text{1}} t)$ 
& LO electric field 
& $E_{\text{2}} = A_{\text{2}} \cos(2\pi f_{\text{2}} t - \phi)$ 
& RF electric field \\

 $E_{\text{total}} \approx 
\big[ A_{\text{1}} + A_{\text{2}} \cos(2\pi \Delta f t + \phi) \big] 
\cos(2\pi f_{\text{1}} t)$ 
& Approximation for $A_{\text{2}}/A_{\text{1}} \ll 1$ & $P(t) = \bar{P}_0 + \kappa A_{\text{2}} \cos(2\pi \Delta f t +  \phi)$ 
& Probe output intensity 
\\

$P_{\text{readout}}(t) = \lvert{P(\Delta f)} \rvert \cos(2\pi \Delta f t + \Delta \phi)$ 
& Detected readout signal 
& $\Omega_{\text{RF}} = \lvert{P(\Delta f)}\rvert / \lvert\kappa \rvert$ 
& RF Rabi frequency\\

\bottomrule
\end{tabular}
\end{table*}

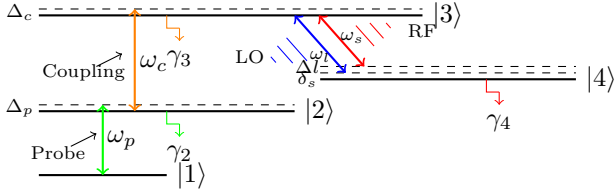
\begin{figure}[h]
\centering
\begin{tikzpicture}[scale=0.85, every node/.style={scale=1},
    level/.style={thick},
    transition/.style={->, >=stealth, thick},
    label/.style={font=\small}
]

\draw[level] (0,0) -- (4/2,0) node[right] {$\ket{1}$};
\draw[dashed] (0,2.2/2) -- (4,2.2/2);
\node at (0.5-0.8,2.1/2) {\scriptsize$\Delta_p$};
\draw[level] (0,2/2) -- (4,2/2) node[right] {$\ket{2}$};
\node at (0.5-0.8,5.1/2) {\scriptsize$\Delta_c$};
\draw[dashed] (0,5.2/2) -- (6,5.2/2);
\draw[level] (0,5/2) -- (6,5/2) node[right] {$\ket{3}$};
\draw[dashed] (4.4,3.4/2) -- (8.4,3.4/2);
\draw[dashed] (4.4,3.2/2) -- (8.4,3.2/2);
\draw[level] (4.4,3/2) -- (8.4,3/2) node[right] {$\ket{4}$};
\node at (6.4-2.2,3.1/2+0.15) {\scriptsize$\Delta l$};
\node at (6.4-2.2,3.3/2-0.15) {\scriptsize$\delta_s$};

\draw[<->, green,thick] (1,0/2) -- (1,2.2/2) ;
\node at (1.3,1.1/2) {$\omega_p$};
\draw[<->, orange,thick] (1.5,2/2) -- (1.5,5.2/2);
\node at (1.8,3.5/2) {$\omega_c$};

\draw [green] (2,2/2) -- (2,0.8) -- (2.2,0.8);
\draw [->,green] (2.2,0.8) -- (2.2,0.6) node[below] {\textcolor{black}{$\gamma_2$}};
\draw [orange] (2,5/2) -- (2,2.3) -- (2.2,2.3);
\draw [->,orange] (2.2,2.3) -- (2.2,2.1) node[below] {\textcolor{black}{$\gamma_3$}};
\draw [red] (7,3/2) -- (7,1.3) -- (7.2,1.3);
\draw [->,red] (7.2,1.3) -- (7.2,1.1) node[below] {\textcolor{black}{$\gamma_4$}};

\coordinate (A) at (4,5/2);
\coordinate (B) at (4.8,3.2/2);
\draw[<->, blue, thick] (A) -- (B);

\def\px{-0.183}
\def\py{-0.081}

\draw[blue] 
  ($ (A)!0.3!(B) + (2*\px,1*\py) $) -- 
  ($ (A)!0.7!(B) + (2*\px,1*\py) $);

\draw[blue] 
  ($ (A)!0.35!(B) + (3*\px,2*\py) $) -- 
  ($ (A)!0.65!(B) + (3*\px,2*\py) $);

\draw[blue] 
  ($ (A)!0.45!(B) + (4*\px,3*\py) $) -- 
  ($ (A)!0.55!(B) + (4*\px,3*\py) $);

\draw[<->, red, thick] (4.38,5/2) -- (5.09,3.4/2);

\def\dx{0.71}
\def\dy{-1.6}
\def\px{0.183}
\def\py{0.0805}

\coordinate (A) at (4.38,5/2);
\coordinate (B) at (5.09,3.4/2);

\draw[red] 
  ($ (A)!0.3!(B) + (2*\px,1*\py) $) -- 
  ($ (A)!0.7!(B) + (2*\px,1*\py) $);

\draw[red] 
  ($ (A)!0.35!(B) + (3*\px,2*\py) $) -- 
  ($ (A)!0.65!(B) + (3*\px,2*\py) $);

\draw[red] 
  ($ (A)!0.55!(B) + (4*\px,3*\py) $) -- 
  ($ (A)!0.45!(B) + (4*\px,3*\py) $);

\node at (6,4.5/2) {\scriptsize RF};
\node at (3.3,3.7/2) {\scriptsize LO};
\node at (4.4,3.7/2) {\scriptsize $\omega_l$};
\node at (4.9,4.3/2) {\scriptsize $\omega_s$};
\node at (0.3,0.7/2) {\scriptsize Probe};

\draw[->] (0.5,0.9/2) -- (0.9,1.3/2);
\node at (0.7,3.2/2) {\scriptsize Coupling};
\draw[->] (0.9,3.4/2) -- (1.3,3.8/2);

\end{tikzpicture}
\caption{Four-level atomic system showing probe, coupling, RF, and LO transitions.}
\label{Fig_0}
\end{figure}

\section{System Model}
\begin{figure}[t] 
\centering
\includegraphics[width=0.4\textwidth]{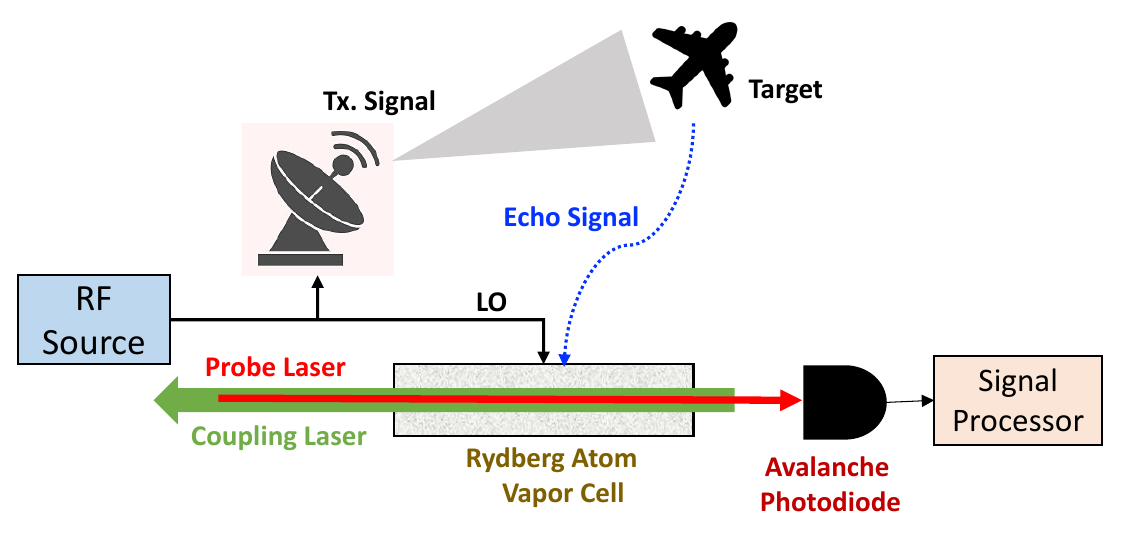}%
\caption{System model of a Rydberg atom-based radar.}
 \label{Fig1}
\end{figure}

A schematic of a Rydberg atom-based radar system is shown in Fig. \ref{Fig1}. An RF source is used to generate a radar signal which is then transmitted by a conventional radar antenna toward an area of interest. Different from conventional radar systems, the Rydberg atom-based radar uses an atomic receiver consisting of Rydberg atoms in a vapor cell in order to improve radar sensitivity. 

In this work, we use the atomic superheterodyne detection principle proposed in \cite{jing2020atomic,gordon2019weak}. For the radar application, the LO signal is derived from the RF source that is used for illuminating the target as shown in Fig.\ \ref{Fig1}. The signal produced by the RF source which is used to illuminate the target as well as the LO is given by
\begin{equation}
    E_1(t) = A_1 \cos(2\pi f_1t),  \quad t\geq 0
    \label{eq2_1}
\end{equation}
The echo signal reflected from the target is given by
\begin{equation}
    E_2(t) = A_2 \cos(2\pi f_2(t-\tau_d) ), \quad  t\geq \tau_d
     \label{eq2_2}
\end{equation}
where $f_2$ is the frequency of the echo signal (different from $f_1$ due to the Doppler shift from the moving target) and $\tau_d$ is the round trip delay given by $\tau_d=2R/c$, where $R$ is the distance of the target from the radar and $c$ denotes the velocity of light. Assuming the target is moving away with a radial velocity of $v$, the frequency of the echo is given by $f_2=(1+2v/c)f_1$.

As shown in Fig. \ref{Fig1}, an atomic superheterodyne detection scheme is used at the receiver where the reflected echo signal and the LO signal interfere in the atomic vapor cell. The total electric field at the Rydberg atom vapor cell can be expressed as
{\scriptsize
\begin{equation}
E(t)=E_1(t)+E_2(t)
= A_1\cos(2\pi f_1 t) + A_2\cos(2\pi f_2 t - \phi), \quad t \geq \tau_d.
\end{equation}
}
where $\phi=2\pi f_2\tau_d$. After some algebra $E(t)$ can be expressed as
{\scriptsize 
\begin{equation}
E(t)=\sqrt{A_1^2 + A_2^2 + 2A_1A_2\cos(2\pi\Delta f t + \phi)}\,\cos(2\pi f_1 t)
= E_s(t)\times E_f(t).
\end{equation}
}
where $\Delta f =f_1-f_2 = 2vf_1/c$ is the intermediate frequency, which contains information about the target's radial velocity. Note that the total electric field inside the Rydberg atom vapor cell $E(t)$ is composed of two components: a fast oscillating field $E_f(t)$ with frequency $f_1$ and a slowly oscillating field $E_s(t)$ with frequency $\Delta f$. The frequency of the control laser is tuned such that the Rydberg atom transition is resonant with $f_1$. Since the intermediate frequency satisfies $\Delta f \ll f_1$, the frequency of $E_s(t)$ is much below the Rydberg transition frequency and the variation in  $E_s(t)$ causes a variation in the intensity of the probe laser. Assuming a strong local oscillator i.e. $A_2\ll A_1 $, $E_s(t)$ can be approximately written as
\begin{align}
   E_s(t) =  A_1 +A_2\cos(2\pi \Delta ft + \phi)\;
    \label{eq2_5}
\end{align}
The variation in $E_s(t)$ leads to a variation in the power of the probe laser \cite{wu2023linear}
\begin{align}
    P(t) = P_0 + C A_2\cos(2\pi \Delta ft + \phi)
     \label{eq2_6}
\end{align}
where $P_0$ is the DC value of the optical power and $C$ (in $\text{W}\cdot\text{m}/\text{V}$) is a constant depending on the laser frequencies and the atomic properties of the Rydberg atom \cite{jing2020atomic,wu2023linear}. 
To derive the expression for the quantity \( C \), we equate \eqref{eq2_6} with RF Rabi frequency equation in Table \ref{tab:rydberg_summary_compact} and utilize a series of interdependent equations and parameter definitions provided in the work \cite{chen2025harnessingrydbergatomicreceivers}. 
Finally, we summarize those expressions, which collectively define the dependencies leading to the final form of \( C \) in Table~\ref{tab:twocol_vertical}.


\begin{table}[h]
\caption{\small List of equations from \cite{chen2025harnessingrydbergatomicreceivers}, required to derive $C$ in \eqref{eq2_6}.}
\label{tab:twocol_vertical}
\centering
\begin{tabular}{|l|l|l|l|}
\hline
\textbf{Quantity} & \textbf{Expression} & \textbf{Quantity} & \textbf{Expression} \\
\hline
$C$ & $C=\kappa\wp_{RF}/\hbar$ 
& $\kappa$ & $\kappa= \alpha \bar{P}_0 \kappa_p$ \\
\hline
$\alpha$ & $\alpha=k_p L C_0 \bar{A}$ 
& $\kappa_p$ & $\kappa_p = \dfrac{\partial \Lambda(\Omega_{\mathrm{LO}}, \Gamma)}{\partial \Omega_{\mathrm{LO}}}$ \\
\hline
$k_p$ & $k_p= 2 \pi / \lambda_p$ 
& $\bar{A}$ & $\bar{A}=\dfrac{\gamma_2 \Omega_p}{\gamma_2^2+2\Omega_p^2}$ \\
\hline
$\Lambda(a,b)$ & $\Lambda(a,b)=\dfrac{b^2}{a^2+b^2}$ 
& $\Gamma$ & $\Gamma=\Omega_p \sqrt{\dfrac{2(\Omega_c^2 + \Omega_p^2)}{2\Omega_p^2 + \gamma_2^2}}$ \\
\hline
\end{tabular}
\end{table}

An avalanche photodiode (APD) is used to convert the oscillating optical power to an electrical signal. The oscillating AC component of the electrical current output from the APD can be expressed as
\begin{equation}
    y(t) = M\mathcal{R} C A_2 \cos(2\pi \Delta ft + \phi) + z(t) \;, \quad t \geq  \tau_d
     \label{eq2_7}
\end{equation}
where $\mathcal{R}$ is the responsivity (in A/W), $M$ is the avalanche multiplication gain and $z(t) \sim \mathcal{N}(0,\sigma_z^2)$ is the additive Gaussian noise. The noise $z(t)$ is composed of two independent noise sources: quantum shot noise arising from the discrete nature of the photons and the electronic thermal noise. The noise variance is given by \cite{keiser2000optical}
\begin{equation}
    \sigma_z^2= 2q(I_0+I_d)M^{2.3} B_e + 4k_BTB_e/R_l \;,
     \label{eq2_8}
\end{equation}
where the first term is the noise variance of the quantum shot noise, and the second term is the variance of the electrical thermal noise. In (\ref{eq2_8}), $q$ denotes the electron charge, $I_0$ is the DC current due to the constant optical power $P_0$, $I_d$ is the dark current, $B_e$ is the electrical bandwidth, $k_B$ is the Boltzmann's constant, $T$ is the operating temperature and $R_l$ is the load resistor.
\section{Performance Analysis}
\subsection{Signal to Noise Ratio}
The signal-to-noise ratio (SNR) for the signal model in (\ref{eq2_7}) can be expressed as
\begin{equation}
    \gamma = \frac{0.5 (M\mathcal{R} C A_2)^2}{2q(I_0+I_d)M^{2.3} B_e + 4k_BTB_e/R_l} \;.
    \label{eq2_9}
\end{equation}
For the echo signal in (\ref{eq2_2}), the average power incident on the Rydberg atom sensor is given by
\begin{equation}
    P_{r} = \frac{A_2^2 A_e}{2Z} \;, 
    \label{eq2_10}
\end{equation}
where $A_e$ is the area of the Rydberg atom sensor and $Z=377 \; \Omega$ is the impedance of the free space. For a target at a distance $R$, the received power is given by
\begin{equation}
    P_r= \frac{P_t G_t  \sigma A_e}{(4\pi)^2R^4} \;,
    \label{eq2_11}
\end{equation}
where $P_t$ is the transmit power, $G_t$ is the transmit antenna gain and $\sigma$ is the radar cross section (RCS). Using (\ref{eq2_10}) and (\ref{eq2_11}) the SNR can be expressed as
\begin{equation}
    \gamma_r = \frac{0.5 (M\mathcal{R} C)^2 \left(\frac{2ZP_tG_t\sigma}{(4\pi)^2R^4} \right) }{2q(I_0+I_d)M^{2.3} B_e + 4k_BTB_e/R_l} \;
    \label{eq2_12}
\end{equation}

The SNR for a standard Radar using a conventional antenna-based receiver is given by
\begin{equation}
    \gamma_s= \frac{P_t G_t \sigma A_s}{(4\pi)^2R^4 k_BT_sB_e} \;,
    \label{eq2_13}
\end{equation}
where $A_s$ is the effective area of the receiver antenna and $T_s$ is the system noise temperature comprising of the noise contributions from the antenna and receiver RF components.



 The Rydberg atom-based receiver provides more flexibility in terms of wideband operation since the operating frequency can be easily tuned by changing the wavelength of the lasers without changing the actual Rydberg atom vapor cell. The operating frequency of the Rydberg atom receiver is independent of the size of the vapor cell whereas the dipole antenna size depends on the carrier frequency since the size of a conventional antenna is of the order $\sim \lambda/2$. We note that the output current signals from the Rydberg atom receiver and the dipole antenna receiver are both corrupted with the additive Gaussian noise; however, the origin of this noise is fundamentally different. The additive noise in the Rydberg atom receiver arises from the avalanche photodetector, whereas the additive noise in the conventional dipole antenna arises from the RF front-end electronics.

\subsection{Target Velocity Estimation}
In order to extract the target information, the Rydberg atom radar performs sampling of the current signal output from the avalanche photodiode (\ref{eq2_7}), for further processing in the digital domain. The sampled discrete time signal is given by
\begin{equation}
y[n] = \alpha \cos\left(2\pi \frac{\Delta f}{f_s} n + \phi\right) + z[n] \;, n=1,\ldots, N \;,
    \label{eq2_14}
\end{equation}
where $f_s$ is the sampling frequency, $\alpha= M\mathcal{R} C A_2$ and $N$ is the total number of samples. Let $\omega = 2\pi \frac{\Delta f}{f_s} $, then the received signal can be expressed as
\begin{equation}
    y[n] = \alpha \cos\left(\omega n + \phi\right) + z[n] \;, n=1,\ldots, N \;
    \label{eq2_15}
\end{equation}
Thus, the velocity of the target can be estimated by estimating the frequency $\omega$ of the output signal from the avalanche photodiode. Frequency estimation of a real sinusoidal signal has been well investigated in the literature. For example, the invariant function approach proposed in  \cite{candan2021frequency} 
 can be used to estimate $\omega$. 

To characterize the fundamental performance limit of any unbiased estimator, the well-known Cram\'{e}r--Rao lower Bound (CRB) is used. The Asymptotic  (ACRB) for frequency estimation is given by \cite{candan2021frequency}
\begin{equation}
    \sigma_{\omega}^{2} \geq \frac{12}{\gamma (N^2-1)N} \; \text{as} \; N \rightarrow \infty.
    \label{eq2_19}
\end{equation}
where $\gamma$ is the SNR of the signal. Since $v= \frac{\omega f_s c}{4 \pi f_1}$, the ACRB for the velocity estimation is given by
\begin{equation}
    \sigma_{v}^{2} \geq \frac{3 f_s^2c^2}{4\pi^2 \gamma (N^2-1)Nf_1^2} \; \text{as} \; N \rightarrow \infty.
    \label{eq2_20}
\end{equation}


\section{Numerical Results}
 The four-level energy structure shown in  is implemented using the sequential transitions $\ket{1} \rightarrow \ket{2} \rightarrow \ket{3} \rightarrow \ket{4}$ in a Cs atom, corresponding to the states 
$6S_{1/2} \rightarrow 6P_{3/2} \rightarrow 30D_{5/2} \rightarrow 31P_{3/2}$. The vapor cell length is $1 ~\mathrm{cm}$ and atomic density at $298 ~\mathrm{K}$ is $4.89 \times 10^{16} m^{-3}$. The transmitted signal frequency is set to  $f_1 = 29.539~\mathrm{GHz}$, which corresponds to the atomic transition frequency $30D_{5/2} \rightarrow 31P_{3/2}$, calculated using the ARC calculator \cite{ROBERTSON2021107814}. 
The probe beam has a power of $20.7~\mu\mathrm{W}$ with a $1/e^2$ beam diameter of $1.7~\mathrm{mm}$. The coupling beam has a power of $17~\mathrm{mW}$ with the same $1/e^2$ beam diameter of $1.7~\mathrm{mm}$. The corresponding Rabi frequencies are calculated as $\Omega_p = 2\pi \times 4.75~\mathrm{MHz}$ and 
$\Omega_c = 2\pi \times 1.66~\mathrm{MHz}$. The amplitude of the transmitted signal is chosen such that the Rabi frequency of the LO field is $\Omega_{LO} = 2\pi \times 0.6~\mathrm{MHz}$. The dipole moments for the relevant transitions are $\wp_{12} = (2.5\,ea_0)^2$ and $\wp_{RF} = 551.35\,ea_0$ 
, where $a_0$ is the Bohr radius. The inverse lifetime of state $\ket{2}$ is $\gamma_{21}/2\pi = 5.2~\mathrm{MHz}$, while the inverse lifetime of the other states are negligible.

\begin{figure}[htp]
\centering
\begin{subfigure}[b]{0.45\textwidth}
    \centering
    \includegraphics[width=0.8\linewidth]{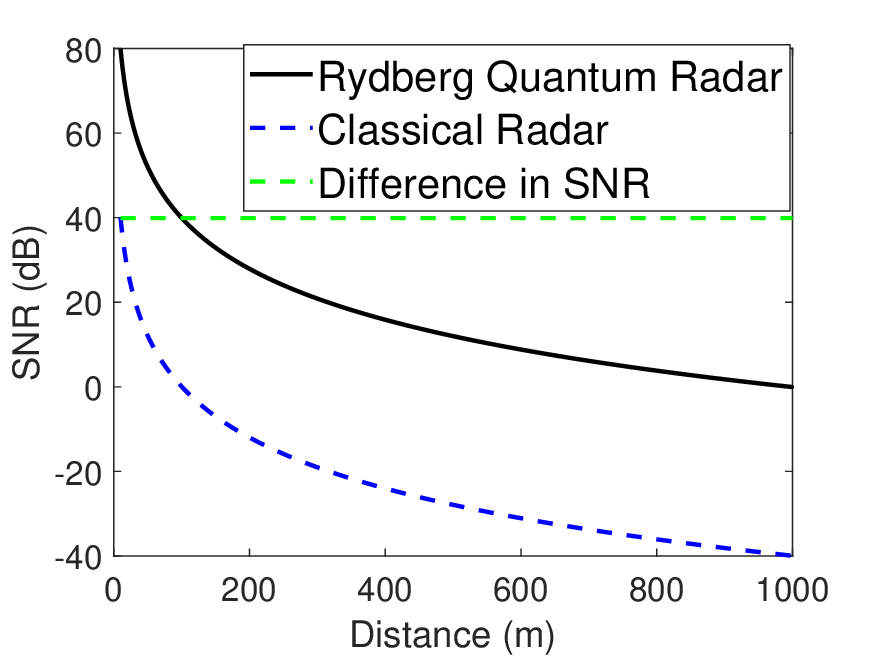}
    \caption{SNR versus distance comparison.}
    \label{Fig_sim1}
\end{subfigure}
\hfill
\begin{subfigure}[b]{0.45\textwidth}
    \centering
    \includegraphics[width=0.8\linewidth]{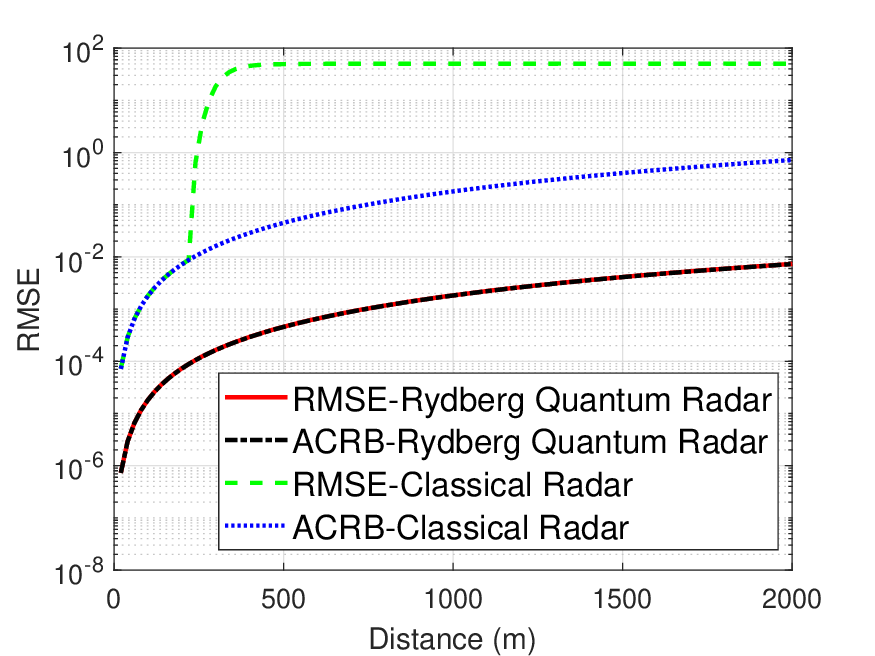}
    \caption{RMSE versus distance comparison.}
    \label{Fig_sim2}
\end{subfigure}
\caption{Performance comparison between Rydberg quantum radar and classical radar: (a) SNR, (b) RMSE of velocity.}
\label{Fig_sim_combined}
\end{figure}

Now, we compare the SNR of the standard radar with that of the Rydberg atom-based radar. The nominal values of the simulation parameters are: $P_t=10$ dBW, $G_t=10$ dB, $\sigma=1 \; \text{m}^2$, $A_e=A_s=1 \; \text{cm}^2$, $T_s=1000$K, $B_e=1$ MHz, $\mathcal{R}=0.6$ A/W, $M=50$, $R_l=1 \; \text{K}\Omega$, $P_0=10\; \mu \text{W}$, $T=300$ K and $I_d=1 \;\text{nA}$. By evaluating the expressions presented in Table~\ref{tab:twocol_vertical}, and using the same parameter values as those employed in \cite{chen2025harnessingrydbergatomicreceivers} for their computation, the value of \( C \) is determined to be \( 6.59 \times 10^{-4} \, \mathrm{W/(V/m)} \). The SNR comparison of the Rydberg atom-based radar and the standard radar is shown in Fig.\ \ref{Fig_sim1}. It can be observed that the SNR of the Rydberg atom radar is significantly higher ($\sim$ 40dB) than that of the conventional radar.


Next, we compare the performance of velocity estimation of the target for the Rydberg radar with that of conventional radar. We assume a target velocity of $v=100$ m/s and a sampling frequency of $f_s=60$ KHz, the number of samples is $N=2048$.  Fig.\ \ref{Fig_sim2} shows the root mean square error (RMSE) of velocity estimation obtained using the invariant function-based frequency estimation \cite{candan2021frequency}for the Rydberg and classical radar. The plots also show the ACRBs for the velocity estimation for benchmarking. It can be observed that the RMSE of the Rydberg radar is lower than that of the classical radar. Furthermore, we observe that when employing the Doppler frequency estimation method described in \cite{candan2021frequency}, the RMSE of the Rydberg quantum radar closely follows the associated ACRB up to a larger target range, approximately 2000 meters. In contrast, the RMSE for the classical radar approaches the ACRB only at shorter distances. These observations highlight the superior performance of the Rydberg quantum radar over classical radar in terms of target velocity estimation accuracy.

\section{Conclusion}
In this work, we proposed a quantum radar system that leverages Rydberg atom-based RF receivers. While the transmitter employed a conventional microwave source, the traditional dipole antenna at the receiver was replaced with a Rydberg atomic RF quantum sensor, offering a fundamentally different detection mechanism. We developed a comprehensive system and signal model for the proposed radar architecture and derived the resulting  SNR achieved with the Rydberg atom-based receiver. This was then compared against the performance of conventional dipole antenna-based receivers. Numerical simulations demonstrated that the Rydberg-based receiver yields a notable improvement in SNR, which translates to enhanced accuracy in target velocity (Doppler frequency) estimation. 





\normalsize

%
%

%
%
\bibliographystyle{IEEEtran}
\bibliography{Rydberg}

\end{document}